\documentclass[preprint1]{aastex}
\usepackage{amsmath}
\slugcomment{}

\newcommand{\be}{\begin{equation}}
\newcommand{\ee}{\end{equation}}

\makeatletter
\renewcommand{\@make@caption@text}[2]{%
  \begin{center}
    \makebox[\textwidth]{\rmfamily#1.\quad#2}
  \end{center}
}%
\makeatother

\shorttitle{Strong Gravitational Lensing Galaxies}
\shortauthors{Melia, Wei \& Wu}

\begin{document}

\title{A Comparison of Cosmological Models Using Strong Gravitational Lensing Galaxies}
\author{Fulvio Melia\altaffilmark{1,2}, Jun-Jie Wei\altaffilmark{1,3}, and Xue-Feng Wu\altaffilmark{1,4,5}}
\altaffiltext{1}{Purple Mountain Observatory, Chinese Academy of Sciences, Nanjing 210008,
China}
\altaffiltext{2}{Department of Astronomy, The Applied Math Program, and Department of Physics,
The University of Arizona, AZ 85721, USA; fmelia@email.arizona.edu.}
\altaffiltext{3}{University of Chinese Academy of Sciences, Beijing 100049, China; jjwei@pmo.ac.cn}
\altaffiltext{4}{Chinese Center for Antarctic Astronomy, Nanjing 210008, China; xfwu@pmo.ac.cn.}
\altaffiltext{5}{Joint Center for Particle, Nuclear Physics and Cosmology, Nanjing
University-Purple Mountain Observatory, Nanjing 210008, China.}

\begin{abstract}
Strongly gravitationally lensed quasar-galaxy systems allow us
to compare competing cosmologies as long as one can be reasonably
sure of the mass distribution within the intervening lens.
In this paper, we assemble a catalog of 69 such systems from
the Sloan Lens ACS and Lens Structure and Dynamics surveys
suitable for this analysis, and carry out a one-on-one comparison
between the standard model, $\Lambda$CDM, and the $R_{\rm h}=ct$
Universe, which has thus far been favored by the application of
model selection tools to other kinds of data. We find that both
models account for the lens observations quite well, though the
precision of these measurements does not appear to be good enough to
favor one model over the other. Part of the reason is the so-called
bulge-halo conspiracy that, on average, results in a baryonic velocity
dispersion within a fraction of the optical effective radius virtually
identical to that expected for the whole luminous-dark matter
distribution modeled as a singular isothermal ellipsoid, though
with some scatter among individual sources. Future work can 
greatly improve the precision of these measurements by focusing 
on lensing systems with galaxies as close as possible to the 
background sources. Given the limitations of doing
precision cosmological testing using the current sample, we 
also carry out  Monte Carlo simulations based on the current lens 
measurements to estimate how large the source catalog would 
have to be in order to rule out either model at a $\sim 99.7\%$ 
confidence level. We find that if the real cosmology is $\Lambda$CDM, 
a sample of $\sim 200$ strong gravitational lenses would be sufficient 
to rule out $R_{\rm h}=ct$ at this level of accuracy, while $\sim 300$ 
strong gravitational lenses would be required to rule out $\Lambda$CDM 
if the real Universe were instead $R_{\rm h}=ct$. The difference
in required sample size reflects the greater number of free parameters 
available to fit the data with $\Lambda$CDM. We point out
that, should the $R_{\rm h}=ct$ Universe eventually emerge as the
correct cosmology, its lack of any free parameters for this kind
of work will provide a remarkably powerful probe of the mass
structure in lensing galaxies, and a means of better 
understanding the origin of the bulge-halo conspiracy.
\end{abstract}

\keywords{cosmology: observations, theory; gravitational lensing: strong;
galaxies: halos, structure; quasars: general}

\section{Introduction}
An interesting new idea started emerging a decade ago (see, e.g., Treu et al.
2006; Grillo et al. 2008; Biesiada et al. 2010; but also see Futamase \&
Yoshida 2001) to use individual lensing galaxies in order to measure
cosmological parameters. In principle, the deflection of quasar light by
the intervening galaxy is known precisely from general relativity as
long as one has a good model for the mass distribution within the lens
(Bartelmann \& Schneider 1999; Refregier 2003). The Einstein radius,
inferred from the deflection angle, then provides a measure of the
angular-size distance, which may be used to discriminate between
competing cosmological models.

The key, of course, is how well we understand the distribution of
matter within the lens, and this appears to be the principal source
of error in this type of measurement. As of today, the observation
of some 70 or so lensing galaxy systems has provided the data
that, in principle, can be used to carry out this kind of study. 
The results thus far are consistent with the standard ($\Lambda$CDM) 
model, though the precision with which model parameters may be 
determined with this appoach is not yet as good as that available in 
other studies, e.g., using Type Ia SNe as standard candles (see, 
e.g., Riess et al. 1998; Perlmutter et al. 1999).

In recent years, the application of model selection tools in one-on-one
comparisons between $\Lambda$CDM and a cosmology we refer to as the
$R_{\rm h}=ct$ Universe (Melia 2007; Melia \& Shevchuk 2012) has shown
that the data actually tend to favor the latter over the former. These
include the use of cosmic chronometers (Melia \& Maier 2013), high-$z$
quasars (Melia 2013), gamma ray bursts (Wei et al. 2013) and, most recently,
the Type Ia SNe themselves (Wei et al. 2014b). The simplest way to
view this cosmology is to start with $\Lambda$CDM and then apply the
additional constraint $p=-\rho/3$ on its total equation of state,
where $p$ and $\rho$ are the total pressure and energy density,
respectively. With these other kinds of data, $R_{\rm h}=ct$ is
favored over $\Lambda$CDM with a likelihood of $\sim 90\%$ versus
only $\sim 10\%$.

The principal goal of this paper is to broaden the comparison between
$R_{\rm h}=ct$ and $\Lambda$CDM by now including strong gravitational
lenses in this study. In \S~2 of this paper, we describe the method,
and then assemble the catalog of suitable lensing systems in \S~3.
We discuss our results in \S~4. We will find that the current strong
lensing sample is not yet large enough to differentiate between these
two competing models, and we show in \S~5 how large the source
catalog needs to be  in order to rule out one or the other expansion
scenarios at a 3-sigma confidence level. We present our conclusions
in \S~6.

\section{Strong Lensing}
The lens model often fitted to the observed images is based on a
singular isothermal ellipsoid (SIE; Ratnatunga et al. 1999), in which
the projected mass distribution (at redshift $z_l$) is elliptical,
with semi-minor axis $\theta_1$ and semi-major axis $\theta_2$. 
In this paper, we will adopt the simpler version, using a singular 
isothermal sphere instead. For generality, we will describe the
approach using semi-major and semi-minor axes, though we
will later set the two angles $\theta_1$ and $\theta_2$ equal
to each other. The source lensed by this system is a quasar 
at redshift $z_s>z_l$. The key expression in strong gravitational 
lensing theory is the lens equation (Schneider et al. 1992), which 
gives the mapping between positions $\beta$ in the source plane 
and $\theta$ in the image plane, according to
\begin{equation}
\beta = \theta -\nabla_\theta \Phi\;.
\end{equation}
The lensing potential of the singular isothermal ellipsoid may
be written (Kormann et al. 1994)
\begin{equation}
\Phi=\theta_{\rm E}\sqrt{(1-\epsilon)\theta_1^2+(1+\epsilon)\theta_2^2}\;,
\end{equation}
where the Einstein radius $\theta_{\rm E}$ is defined below in terms of
the (one-dimensional) velocity dispersion, $\sigma_v$,  in the lensing galaxy, 
and the angular diameter distances, $D_A(z_l,z_s)$ and $D_A(0,z_s)$, 
between lens and source and between source and observer, respectively.
The `ellipticity' $\epsilon$ is related to the eccentricity $e$ of the
critical line by
\begin{equation}
e=\sqrt{(1-\epsilon)/(1+\epsilon)}\;.
\end{equation}
In addition, the semi-major and semi-minor axes are related to
the Einstein radius
\begin{equation}
\theta_{\rm E}\equiv 4\pi\left({\sigma_v\over c}\right)^2{\mathcal{D}}\;,
\end{equation}
where
\begin{equation}
{\mathcal{D}}\equiv {D_A(z_l,z_s)\over D_A(0,z_s)}\;,
\end{equation}
via the relations
\begin{eqnarray}
\theta_1&=&\theta_{\rm E}\sqrt{1-\epsilon}\nonumber \\
\theta_2&=&\theta_{\rm E}\sqrt{1+\epsilon}\;.
\end{eqnarray}
As noted, earlier, we will here consider the simpler case of a single 
isothermal sphere (SIS), for which $\theta_1=\theta_2$.

In principle, equation~(4) can be used to test cosmological models in a
rather unique way because, unlike other kinds of comparisons that rely
on the optimization of the Hubble constant $H_0$, this particular analysis
is completely independent of $H_0$.\footnote{One can in fact
determine $H_0$ directly using strong gravitational lensing, but only
when time delays are measured between the various images of a given
source (see, e.g., Paraficz \& Hjorth 2009; Suyu et al. 2013; Wei
et al. 2014a).} Nonetheless, even though knowledge of $H_0$ is 
not necessary for this type of test, fits to the data do depend on 
the reliability of lens modelling (e.g., via the assumption of a singular 
isothermal sphere, or a singular isothermal ellipsoid) and the 
measurement of the velocity dispersion.

When using these expressions, $\sigma_v$ (the total velocity dispersion
of stellar plus dark matter) cannot be obtained directly from the
surface-brightness weighted average of the line-of-sight velocity dispersion
that is actually measured. In practice, the central velocity dispersion $\sigma_0$
is estimated from the {\it stellar} velocity dispersion within $R_e/8$, where
$R_e$ is the optical effective radius (see, e.g., Treu et al. 2006; Grillo
et al. 2008), and is then used to represent the velocity dispersion
$\sigma_{SIS}$ for the corresponding singular isothermal sphere or
ellpisoid (for the total mass present).
This works rather well because inside one effective radius, massive
elliptical galaxies are kinematically indistinguishable from an
isothermal ellipsoid (Koopmans et al. 2009), which is quite remarkable
considering the fact that $\sigma_{SIS}$ and $\sigma_0$ need not be
the same. One reason is that dark matter halos appear to be dynamically
hotter than the luminous stars (based on X-ray observations), so the
former must necessarily have a greater velocity dispersion than the
latter (White \& Davis 1996).

Still, when one introduces the SIS equivalent value $\sigma_{SIS}$ obtained
from modelling the lens as a singular isothermal sphere, these two
measures of velocity dispersion agree very closely. Treu et al. (2006) used
the large and homogeneously selected sample of lenses identified by the Sloan
Lenses ACS Survey (SLACS; Bolton et al. 2005, 2006) to study in detail the
degree of homogeneity of the early-type galaxies by measuring the ratio
between stellar velocity dispersion and $\sigma_{SIS}$ that best fits the
geometry of the corresponding multiple images. They found that the ratio
${\sigma_0/ \sigma_{SIS}}$ is very close to unity; specifically, they
inferred a sample average value $\langle {\sigma_0/ \sigma_{SIS}}\rangle=
1.010\pm 0.017$, with a relatively small scatter of $\sim 0.06$. Similarly,
van de Ven et al. (2003) examined this ratio for a range of anisotropy
parameters and found that $0.96<{\sigma_0/ \sigma_{SIS}}<1.08$. The
conclusion from such studies is that on average the approximation
$\sigma_v=\sigma_{SIS}\approx\sigma_0$ works surprisingly well, due perhaps
to some as yet unknown mechanism that couples stellar and dark mass,
sometimes referred to as a bulge-halo `conspiracy'.  A possible resolution 
of this parity may be that since the NFW (Navarro et al. 1997) and observed 
stellar mass profiles are nearly isothermal, the more concentrated
mass profiles for the baryon component than for dark matter may
simply be a consequence of dissipative star formation. The observation
of $\sigma_0/\sigma_{SIS}\sim 1$ may therefore not  be so 
mysterious. Nonetheless, significant departures from this are seen 
in individual cases, so one cannot ignore the scatter in any discussion 
concerning the propagated measurement error for ${\mathcal{D}}_{obs}$.

In this paper, we will follow Cao et al. (2012), and put
\begin{equation}
\theta_{\rm E}\equiv 4\pi\left({\sigma_{SIS}\over c}\right)^2{\mathcal{D}}\;,
\end{equation}
with
\begin{equation}
\sigma_{SIS}\equiv f_{SIS}\,\sigma_0\;.
\end{equation}
We will keep $f_{SIS}$ as a free parameter to be optimized in the
fits, since it mimics at least several effects that apparently give
rise to the observed scatter in the individually measured ratio
$\sigma_0/\sigma_{SIS}$ for each system. These include: (1) systematic
errors in the rms deviation of $\sigma_{SIS}$ from $\sigma_0$; (2)
an rms error associated with the assumption that the SIS model
allows the translation from observed image separation to $\theta_{\rm E}$;
and (3) a softened isothermal sphere potential, which tends to decrease
the typical image separations (Narayan \& Bartelmann 1996). In the
analysis we describe below, we will adopt a dispersion $\sigma_f=0.06\,f_{SIS}$,
based on the rms scatter of $\sim 6\%$ found from
the work of Treu et al. (2006) and van de Ven et al. (2003). Note,
however, that $\sigma_f$ may be as big as $\sim 0.2$ according to
Cao et al. (2012), though one might have expected such a large value
to have emerged directly from the aforementioned survey by Treu et
al. (2006).

The overall uncertainty associated with ${\mathcal{D}_{obs}}$ (calculated
from equation~7) is estimated through the propagation equation involving
errors in $\theta_{\rm E}$, $\sigma_0$, and $f_{SIS}$. According to
Grillo et al. (2008), the error in measuring the Einstein radius
$\theta_{\rm E}$ is $\sim 5\%$, so we will assume a dispersion
$\sigma_{\theta_{\rm E}}=0.05\,\theta_{\rm E}$ for this quantity. In principle, any
optimization of the model parameters (and $f_{SIS}$) carried out while
fitting the data should also include the dispersion $\sigma_z$ in the
measured redshifts $z_l$ and $z_s$ (since the theoretical values
$D_A(z_l,z_s)$ and $D_A(0,z_s)$ directly depend on these).
However, a careful analysis
of SDSS quasar spectra shows that $\sigma_z/(1+z)\sim 10^{-4}$ over a
broad range of redshifts (Hewett \& Wild 2010). This error is so small
compared to the other three uncertainties that we will ignore it.
So in total, we will calculate the dispersion $\sigma_{\mathcal{D}}$
in ${\mathcal{D}_{obs}}$ using the expression
\begin{equation}
\sigma_{\mathcal{D}}={\mathcal{D}_{obs}}\left[\left({\sigma_{\theta_{\rm E}}\over
\theta_{\rm E}}\right)^2+4\left({\sigma_{\sigma_{0}}\over \sigma_0}\right)^2
+4\left({\sigma_f\over f_{SIS}}\right)^2\right]^{1/2}\;.
\end{equation}
Note that since the uncertainty in $\sigma_0$ also appears to be $\sim 5\%$
(Grillo et al. 2008), the average dispersion in the measured value of $\mathcal{D}$
is expected to be $\sigma_{\mathcal{D}}\sim 0.16\,{\mathcal{D}_{obs}}$. 

Now, in principle, only the range $0\le {\mathcal{D}}\le 1$ is physically 
meaningful, but such a value of $\sigma_{\mathcal{D}}$ can result in at least 
some measurements ${\mathcal{D}_{obs}}>1.0$. A quick inspection of 
equation~(5) shows that the measurements most at risk for this type of 
outcome involve lenses much closer to the observer than the source. As 
we shall see below, several of the sources in our complete sample have 
${\mathcal{D}_{obs}}>1.0$. Though unrealistic, such values are 
consistent with the quoted error, so we will include them in our analysis.
But to demonstrate their negative impact on the optimization of the model fits,
we will also carry out the analysis for a reduced sample omitting these
sources. As the measurements become more precise, and the sample 
of strong lensing sources grows, it may be possible to avoid systems
with ${\mathcal{D}_{obs}}>1.0$ altogether.

From a theoretical standpoint, one must assume a cosmological
model in order to calculate the angular diameter distances $D_A(z_l,z_s)$
and $D_A(0,z_s)$, once the redshifts $z_l$ and $z_s$ for a particular
lensing system are known. In $\Lambda$CDM, this distance depends
on several parameters, including $H_0$ and the mass fractions
$\Omega_{\rm m} \equiv \rho_{\rm m}/\rho_{\rm c}$, $\Omega_{\rm r}\equiv
\rho_{\rm r}/\rho_{\rm c}$, and $\Omega_{\rm de}\equiv \rho_{\rm de}/
\rho_{\rm c}$, defined in terms of the current matter ($\rho_{\rm m}$),
radiation ($\rho_{\rm r}$), and dark energy ($\rho_{\rm de}$) densities,
and the critical density $\rho_{\rm c}\equiv 3c^2H_0^2/8\pi G$.
Assuming zero spatial curvature, so that $\Omega_{\rm m}+\Omega_{\rm r}
+\Omega_{\rm de}=1$, the angular diameter distance between redshifts
$z_1$ and $z_2$ ($>z_1$) is given by the expression
\begin{equation}
D_A^{\Lambda{\rm CDM}}(z_1,z_2)={c\over H_0}{1\over (1+z_2)}\int_{z_1}^{z_2}
\left[\Omega_{\rm m}(1+z)^3+\Omega_{\rm r}(1+z)^4+\Omega_{\rm de}
(1+z)^{3(1+w_{\rm de})}\right]^{-1/2}\;dz\;,
\end{equation}
where $p_{\rm de}=w_{\rm de}\rho_{\rm de}$ is the dark-energy equation
of state. As noted earlier, $H_0$ cancels out when we divide
$D_A^{\Lambda{\rm CDM}}(z_l,z_s)$ by $D_A^{\Lambda{\rm CDM}}(0,z_s)$
to form the ratio $\mathcal{D}_{\Lambda{\rm CDM}}$,
so the essential remaining parameters in flat $\Lambda$CDM are
$\Omega_{\rm m}$ and $w_{\rm de}$. If we further assume that
dark energy is a cosmological constant with $w_{\rm de}=-1$,
then only the parameter $\Omega_{\rm m}$ is available to fit
the data.

In the $R_{\rm h}=ct$ Universe (Melia 2007; Melia \& Shevchuk 2012),
the angular diameter distance depends only on $H_0$, but since
here too the Hubble constant cancels out in the ratio $\mathcal{D}_{R_{\rm h}=ct}$,
there are actually no free parameters left for fitting the gravitational
lensing data. In this cosmology,
\begin{equation}
D_A^{R_{\rm h}=ct}(z_1,z_2)={c\over H_0}{1\over (1+z_2)}
\ln\left({1+z_2\over 1+z_1}\right)\;.
\end{equation}

\section{Strong Gravitational Lensing Systems}
 Our sample is drawn from a compilation of 69 strong lensing systems
(listed in Table I), with
good spectroscopic measurements of the central velocity dispersion, using
the SLACS ({\it Sloan Lens ACS}) Survey (first introduced by Bolton et al.
2006; Treu et al. 2006; and Koopmans et al. 2006), and the LSD ({\it Lenses
Structure and Dynamics}) Survey (see, e.g., Bolton et al. 2008; Newton
et al. 2011). Some original contributions to these data sets may be found
in Young et al. (1980), Huchra et al. (1985), Leh\'ar et al. (1993),
Fassnacht et al. (1996), Tonry et al. (1998), Koopmans \& Treu (2002, 2003),
and Treu and Koopmans (2004). The velocity dispersion $\sigma_0$ and its
uncertainty (with the aforementioned average value of $\sim 5\%$) were
obtained from the {\it Sloan Digital Sky Survey Database}. The SLACS and
LENS surveys complement each other rather well, with the former comprised
primarily of lens galaxies at redshift up to $\sim 0.3$, while the latter
includes systems beyond $z\sim 0.5$.

It has already been noted before (see, e.g., Biesiada et al.
2010, 2011; Cao et al. 2012) that some of these lenses produce 2 images,
while others produce 4. Both 2-image and 4-image lens systems are 
usually affected by external shear. This effect degenerates with the 
ellipticity of the SIE component, which introduces some uncertainty 
in estimating the Einstein radius of a given lens. For 2-image
systems this can be more problematic due to a lack of observational 
constraints. On the other hand, 4-image systems are better constrained
observationally, so both their ellipticity and external shear may be
determined for a more accurate measurement of the Einstein radius.
To gauge whether there are any systematic effects associated with
one category or the other, we will here track the results using both 
sets of lens system.

There is an additional drawback to the measurement of $\mathcal{D}$ 
with strong gravitational lenses that we must carefully consider here. By 
its very definition, $\mathcal{D}$ is confined
to a very compact range of values $(0,1)$, regardless of the lens ($z_l$)
and quasar ($z_s$) redshifts. In addition, there is no monotonic progression
from low to high values of $\mathcal{D}$ as the sequence of gravitational
lenses approaches or recedes from us, since the sources may lie anywhere
beyond them. Fortunately, $\mathcal{D}$ does not depend on $H_0$, so a
comparison between theoretical values of this ratio and $\mathcal{D}_{\rm obs}$
is not inhibited by any uncertainty in the expansion rate itself.
However, the tight range in $\mathcal{D}$ and its lack of correlation
with $z$ make it difficult to optimize parameters such as $\Omega_{\rm m}$,
which produce only slight changes in $\mathcal{D}_{\Lambda{\rm CDM}}$ even
when they increase by a factor of 2 or more. In this paper, we will
therefore compare how well $R_{\rm h}=ct$ fits the data with several
specific variations of $\Lambda$CDM, though always assuming a flat spatial
curvature and a cosmological constant with $w_{\rm de}=-1$. The most prominent 

\begin{deluxetable}{lllccclccl}
\tablewidth{520pt}
\tabletypesize{\footnotesize}
\tablecaption{Strong Gravitational Lensing Systems}\tablenum{1}
\tablehead{\colhead{Galaxy}&\colhead{$z_l$}&\colhead{$z_s$}&\colhead{$\theta_{\rm E}$}&\colhead{$\sigma_0$}&
\colhead{$\mathcal{D}_{\rm obs}$}&\colhead{$\sigma_{\mathcal{D}}$}&
\colhead{${\mathcal{D}_{R_{\rm h}=ct}}$}&\colhead{${\mathcal{D}_{\Lambda{\rm CDM}}}$}& \colhead{Refs.} \\
&&&({\rm arcsec})&(${\rm km}\;{\rm s}^{-1})$&$f_{\rm SIS}=1.02$&&& $(\Omega_{\rm m}=0.27)$&
} \startdata
\sidehead{\centerline{\qquad\quad Systems with Two Images}}
{\rm SDSS J}0037-0942 &  0.1955 &  0.6322  &  1.47  &  282$\pm$11  &  0.617  &  0.094  &  0.636  &  0.656  &  1--9 \\
{\rm SDSS J}0216-0813 &  0.3317 &  0.5235  &  1.15  &  349$\pm$24  &  0.316  &  0.060  &  0.320  &  0.336  &  1--9 \\
{\rm SDSS J}0737+3216 &  0.3223 &  0.5812  &  1.03  &  326$\pm$16  &  0.323  &  0.053  &  0.390  &  0.409  &  1--9 \\
{\rm SDSS J}0912+0029 &  0.1642 &  0.3240  &  1.61  &  325$\pm$12  &  0.509  &  0.076  &  0.458  &  0.474  &  1--9 \\
{\rm SDSS J}1250+0523 &  0.2318 &  0.7950  &  1.15  &  274$\pm$15  &  0.511  &  0.087  &  0.644  &  0.665  &  1--9 \\
{\rm SDSS J}1630+4520 &  0.2479 &  0.7933  &  1.81  &  279$\pm$17  &  0.776  &  0.138  &  0.621  &  0.642  &  1--9 \\
{\rm SDSS J}2300+0022 &  0.2285 &  0.4635  &  1.25  &  305$\pm$19  &  0.448  &  0.081  &  0.460  &  0.479  &  1--9 \\
{\rm SDSS J}2303+1422 &  0.1553 &  0.5170  &  1.64  &  271$\pm$16  &  0.745  &  0.131  &  0.654  &  0.673  &  1--9 \\
{\rm CFRS}03.1077     &  0.9380 &  2.9410  &  1.24  &  251$\pm$19  &  0.657  &  0.131  &  0.518  &  0.506  &  1--9 \\
{\rm HST} 15433       &  0.4970 &  2.0920  &  0.36  &  116$\pm$10  &  0.893  &  0.193  &  0.643  &  0.652  &  1--9 \\
{\rm MG}2016          &  1.004 &  3.263  &  1.56  &  328$\pm$32  &  0.484  &  0.113  &  0.521  &  0.504  &  1--9 \\
{\rm SDSS J}0044+0113 &  0.1196 &  0.1965  &  0.79  &  266$\pm$13  &  0.373  &  0.061  &  0.370  &  0.381  &  10,11 \\
{\rm SDSS J}0330-0020 &  0.3507 &  1.0709  &  1.10  &  212$\pm$21  &  0.817  &  0.194  &  0.587  &  0.608  &  10,11 \\
{\rm SDSS J}0935-0003 &  0.3475 &  0.4670  &  0.87  &  396$\pm$35  &  0.185  &  0.041  &  0.222  &  0.234  &  10,11 \\
{\rm SDSS J}0955+0101 &  0.1109 &  0.3159  &  0.91  &  192$\pm$13  &  0.824  &  0.155  &  0.617  &  0.632  &  10,11 \\
{\rm SDSS J}0959+4416 &  0.2369 &  0.5315  &  0.96  &  244$\pm$19  &  0.538  &  0.109  &  0.501  &  0.521  &  10 \\
{\rm SDSS J}1112+0826 &  0.2730 &  0.6295  &  1.48  &  320$\pm$20  &  0.486  &  0.088  &  0.506  &  0.527  &  10,11 \\
{\rm SDSS J}1142+1001 &  0.2218 &  0.5039  &  0.98  &  221$\pm$22  &  0.670  &  0.159  &  0.509  &  0.529  &  10,11 \\
{\rm SDSS J}1143-0144 &  0.1060 &  0.4019  &  1.68  &  269$\pm$13  &  0.775  &  0.126  &  0.702  &  0.718  &  10 \\
{\rm SDSS J}1204+0358 &  0.1644 &  0.6307  &  1.31  &  267$\pm$17  &  0.613  &  0.112  &  0.689  &  0.708  &  10,11 \\
{\rm SDSS J}1205+4910 &  0.2150 &  0.4808  &  1.22  &  281$\pm$14  &  0.516  &  0.084  &  0.504  &  0.524  &  10 \\
{\rm SDSS J}1213+6708 &  0.1229 &  0.6402  &  1.42  &  292$\pm$15  &  0.556  &  0.092  &  0.766  &  0.783  &  10,11 \\
{\rm SDSS J}1403+0006 &  0.1888 &  0.4730  &  0.83  &  213$\pm$17  &  0.611  &  0.126  &  0.553  &  0.573  &  10 \\
{\rm SDSS J}1436-0000 &  0.2852 &  0.8049  &  1.12  &  224$\pm$17  &  0.745  &  0.149  &  0.575  &  0.597  &  10,11\\
{\rm SDSS J}1443-0304 &  0.1338 &  0.4187  &  0.81  &  209$\pm$11  &  0.619  &  0.104  &  0.641  &  0.658  &  10,11 \\
{\rm SDSS J}1451-0239 &  0.1254 &  0.5203  &  1.04  &  223$\pm$14  &  0.698  &  0.126  &  0.718  &  0.735  &  10,11 \\
{\rm SDSS J}1525+3327 &  0.3583 &  0.7173  &  1.31  &  264$\pm$26  &  0.627  &  0.148  &  0.434  &  0.454  &  10,11 \\
{\rm SDSS J}1531-0105 &  0.1596 &  0.7439  &  1.71  &  279$\pm$14  &  0.733  &  0.120  &  0.734  &  0.753  &  10,11 \\
{\rm SDSS J}1538+5817 &  0.1428 &  0.5312  &  1.00  &  189$\pm$12  &  0.934  &  0.170  &  0.687  &  0.705  &  10,11 \\
{\rm SDSS J}1621+3931 &  0.2449 &  0.6021  &  1.29  &  236$\pm$20  &  0.773  &  0.165  &  0.535  &  0.556  &  10,11 \\
{\rm MG}1549+3047     &  0.11 &  1.17  &  1.15  &  227$\pm$18  &  0.745  &  0.153  &  0.865  &  0.878  &  12 \\
{\rm CY}2201-3201       &  0.32 &  3.90  &  0.41  &  130$\pm$20  &  0.810  &  0.270  &  0.825  &  0.764  &  2,4,5 \\
{\rm SDSS J}1432+6317 & 0.1230 & 0.6643  &  1.26  &  199$\pm$10  &  1.062  &  0.174  &  0.772  &  0.749  &  10,11 \\
{\rm SDSS J}2238-0754  & 0.1371 & 0.7126  &  1.27  &  198$\pm$11  &  1.081  &  0.185  &  0.761  &  0.736  &   10,11 \\
{\rm Q}0957+561           & 0.36 & 3.90  &  1.41  &  167$\pm$10  &  1.077  &  0.190  &  0.650  &  0.599  &   13 \\
\cline{1-10}
\sidehead{\centerline{\qquad\quad Systems with More than Two Images}}
{\rm SDSS J}0956+5100 &  0.2405 &  0.4700  &  1.32  &  318$\pm$17  &  0.436  &  0.073  &  0.441  &  0.459  &  1--9 \\
{\rm SDSS J}0959+0410 &  0.1260 &  0.5349  &  1.00  &  229$\pm$13  &  0.636  &  0.110  &  0.723  &  0.740  &  1--9 \\
{\rm SDSS J}1330-0148 &  0.0808 &  0.7115  &  0.85  &  195$\pm$10  &  0.746  &  0.124  &  0.855  &  0.868  &  1--9 \\
{\rm SDSS J}1402+6321 &  0.2046 &  0.4814  &  1.39  &  290$\pm$16  &  0.552  &  0.094  &  0.526  &  0.546  &  1--9 \\
{\rm SDSS J}1420+6019 &  0.0629 &  0.5352  &  1.04  &  206$\pm$5   &  0.818  &  0.113  &  0.858  &  0.869  &  1--9 \\
{\rm SDSS J}1627-0053 &  0.2076 &  0.5241  &  1.21  &  295$\pm$13  &  0.464  &  0.073  &  0.552  &  0.573  &  1--9 \\
{\rm SDSS J}2321-0939 &  0.0819 &  0.5324  &  1.57  &  245$\pm$7   &  0.873  &  0.124  &  0.816  &  0.829  &  1--9 \\
{\rm Q}0047-2808      &  0.4850 &  3.5950  &  1.34  &  229$\pm$15  &  0.853  &  0.157  &  0.741  &  0.738  &  1--9 \\
{\rm HST} 14176       &  0.8100 &  3.3990  &  1.41  &  224$\pm$15  &  0.938  &  0.175  &  0.599  &  0.587  &  1--9 \\
{\rm SDSS J}0029-0055 &  0.2270 &  0.9313  &  0.96  &  229$\pm$18  &  0.611  &  0.125  &  0.689  &  0.710  &  10,11 \\
{\rm SDSS J}0109+1500 &  0.2939 &  0.5248  &  0.69  &  251$\pm$19  &  0.366  &  0.073  &  0.389  &  0.407  &  10 \\
{\rm SDSS J}0728+3835 &  0.2058 &  0.6877  &  1.25  &  214$\pm$11  &  0.911  &  0.151  &  0.642  &  0.663  &  10,11 \\
{\rm SDSS J}0822+2652 &  0.2414 &  0.5941  &  1.17  &  259$\pm$15  &  0.582  &  0.101  &  0.536  &  0.557  &  10,11 \\
{\rm SDSS J}0841-3824 &  0.1159 &  0.6567  &  1.41  &  225$\pm$11  &  0.930  &  0.151  &  0.783  &  0.799  &  10,11 \\
{\rm SDSS J}0936+0913 &  0.1897 &  0.5880  &  1.09  &  243$\pm$12  &  0.616  &  0.101  &  0.624  &  0.645  &  10 \\
{\rm SDSS J}0946+1006 &  0.2219 &  0.6085  &  1.38  &  263$\pm$21  &  0.666  &  0.137  &  0.609  &  0.599  &  10,11 \\
{\rm SDSS J}1016+3859 &  0.1679 &  0.4349  &  1.09  &  247$\pm$13  &  0.596  &  0.100  &  0.570  &  0.589  &  10 \\
{\rm SDSS J}1020+1122 &  0.2822 &  0.5530  &  1.20  &  282$\pm$18  &  0.504  &  0.092  &  0.435  &  0.455  &  10 \\
{\rm SDSS J}1023+4230 &  0.1912 &  0.6960  &  1.41  &  242$\pm$15  &  0.804  &  0.144  &  0.669  &  0.808  &  10,11 \\
{\rm SDSS J}1029+0420 &  0.1045 &  0.6154  &  1.01  &  210$\pm$11  &  0.764  &  0.128  &  0.793  &  0.808  &  10 \\
{\rm SDSS J}1032+5322 &  0.1334 &  0.3290  &  1.03  &  296$\pm$15  &  0.392  &  0.065  &  0.560  &  0.576  &  10 \\
{\rm SDSS J}1103+5322 &  0.1582 &  0.7353  &  1.02  &  196$\pm$12  &  0.886  &  0.158  &  0.734  &  0.752  &  10,11 \\
{\rm SDSS J}1106+5228 &  0.0955 &  0.4069  &  1.23  &  262$\pm$13  &  0.598  &  0.098  &  0.733  &  0.748  &  10,11 \\
{\rm SDSS J}1134+6027 &  0.1528 &  0.4742  &  1.10  &  239$\pm$12  &  0.643  &  0.106  &  0.634  &  0.652  &  10 \\
{\rm SDSS J}1153+4612 &  0.1797 &  0.8751  &  1.05  &  226$\pm$15  &  0.686  &  0.127  &  0.737  &  0.756  &  10 \\
{\rm SDSS J}1416+5136 &  0.2987 &  0.8111  &  1.37  &  240$\pm$25  &  0.794  &  0.195  &  0.560  &  0.582  &  10,11 \\
{\rm SDSS J}1430+4105 &  0.2850 &  0.5753  &  1.52  &  322$\pm$32  &  0.489  &  0.116  &  0.448  &  0.468  &  10 \\
{\rm SDSS J}1636+4707 &  0.2282 &  0.6745  &  1.09  &  231$\pm$15  &  0.682  &  0.125  &  0.601  &  0.623  &  10 \\
{\rm PG}1115+080      &  0.3100 &  1.7200  &  1.21  &  281$\pm$25  &  0.511  &  0.113  &  0.730  &  0.745  &  14 \\
{\rm Q}2237+030       &  0.04 &   1.169   &  0.91  &   215$\pm$30  &  0.657  &  0.202  &  0.949  &  0.940  &  15 \\
{\rm B}1608+656        &  0.63 &   1.39    & 1.13   &   247$\pm$35  &  0.618  &  0.193  &  0.439  &  0.386  &  16 \\
{\rm SDSS J}0252+0039 & 0.2803  &  0.9818  &  1.04  & 164$\pm$12  &  1.290  & 0.253 & 0.639 & 0.599 & 10 \\
{\rm SDSS J}0405-0455  & 0.0753  &  0.8098  &  0.80  & 160$\pm$8  &  1.043  & 0.171  & 0.878 & 0.861 & 10 \\
{\rm SDSS J}2341+0000 & 0.186 & 0.807 & 1.44  & 207$\pm$13  &  1.122  & 0.203 & 0.712 & 0.681 & 10,11 \\
\enddata
\tablenotetext{References:\hskip0.2in} {(1) Treu \& Kooopmans (2002); (2) Koopmans \& Treu (2002);
(3) Treu \& Koopmans (2003); (4) Koopmans \& Treu (2003); (5) Treu \& Koopmans (2004); (6) Treu et al.
(2006); (7) Kooopmans et al. (2006); (8) Grillo et al. (2008); (9) Biesiada, Pi\'orkowska
\& Malec (2010); (10) Bolton et al. (2008); (11) Newton et al. (2011); (12) Leh\'ar et al. (1993);
(13) Young et al. (1980); (14) Tonry (1998); (15) Huchra et al. (1985); (16) Fassnacht et al. (1986)
}
\end{deluxetable}

\noindent comparison will be between $R_{\rm h}=ct$ and the concordance
model (with $\Omega_{\rm m}=0.27$), though we will also consider other
values of $\Omega_{\rm m}$, including the Einstein-de Sitter (E-deS) model
with $\Omega_{\rm m}=1$.

Related to the possible difficulty in using sources with $\mathcal{D}_{\rm obs} >1$
is the fact that the uncertainty $\sigma_{\mathcal{D}}$ in $\mathcal{D}_{\rm obs}$ 
carries significantly more weight when $\mathcal{D}_{\rm obs}\gtrsim 0.6$ than 
elsewhere in its permitted range, because here big changes in $z_s$ produce 
only very slight modifications to $\mathcal{D}_{\Lambda{\rm CDM}}$ and 
$\mathcal{D}_{R_{\rm h}=ct}$. One therefore sees an increasing scatter 
among the observed values of $\mathcal{D}_{\rm obs}$, as shown in 
figures~1 and 2. In order to fully understand the impact of all of these
issues, we will analyze the quality of the theoretical fit for both the full 
sample and a reduced sample with $\mathcal{D}_{\rm obs}<1$, and
in each case also a sub-sample of 2-image systems only. Figures~1 and 2
show the results of analyzing all lensing systems with 2 images only. The
figures corresponding to other sample selection criteria are very similar.

\begin{figure}[hp]
\centerline{\includegraphics[angle=0,scale=0.7]{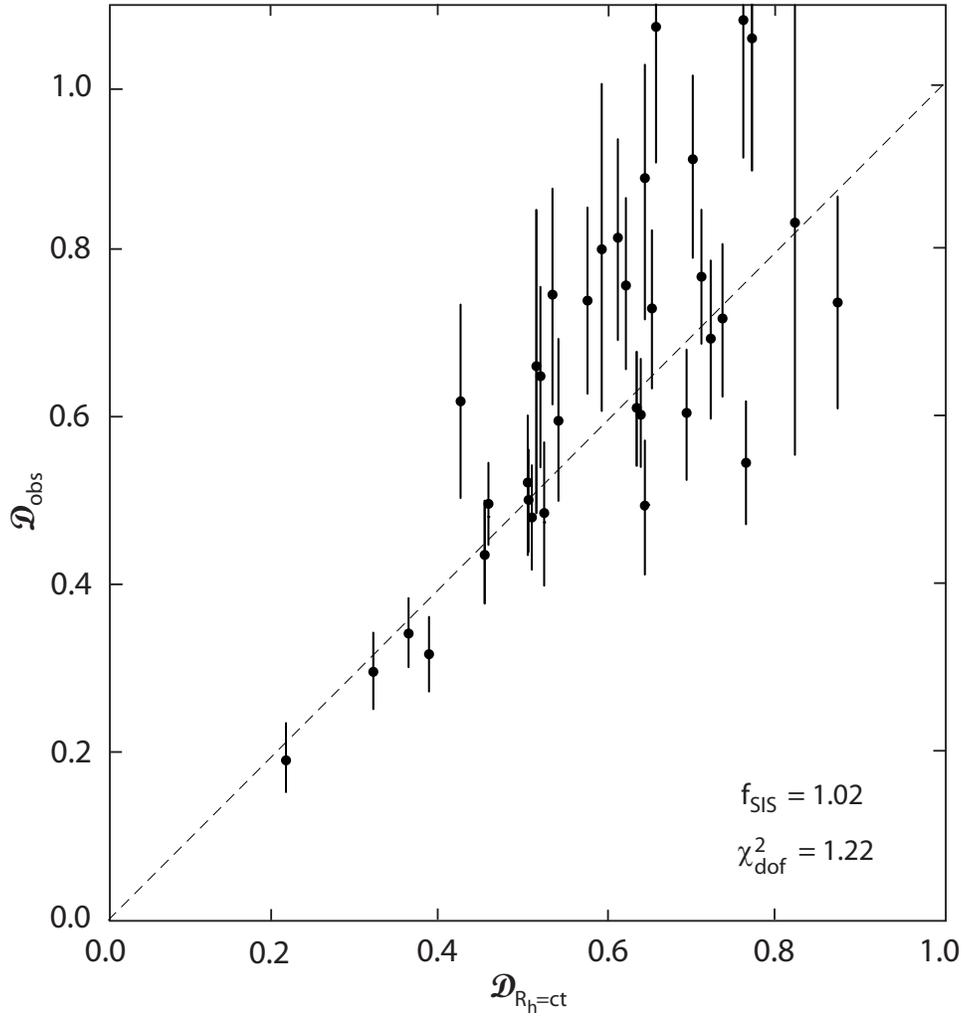}}
\vskip-0.2in
\caption{Observed value of $\mathcal{D}$ versus that predicted in the
$R_{\rm h}=ct$ Universe for all lensing systems with 2 images only. 
A perfect fit would correspond to the dashed
diagonal line. The optimized value of $f_{\rm SIS}$ in this case
is $1.02$, and the reduced $\chi^2$ is $1.22$, with $34-1=33$ degrees
of freedom.}
\end{figure}

\begin{figure}[hp]
\centerline{\includegraphics[angle=0,scale=0.7]{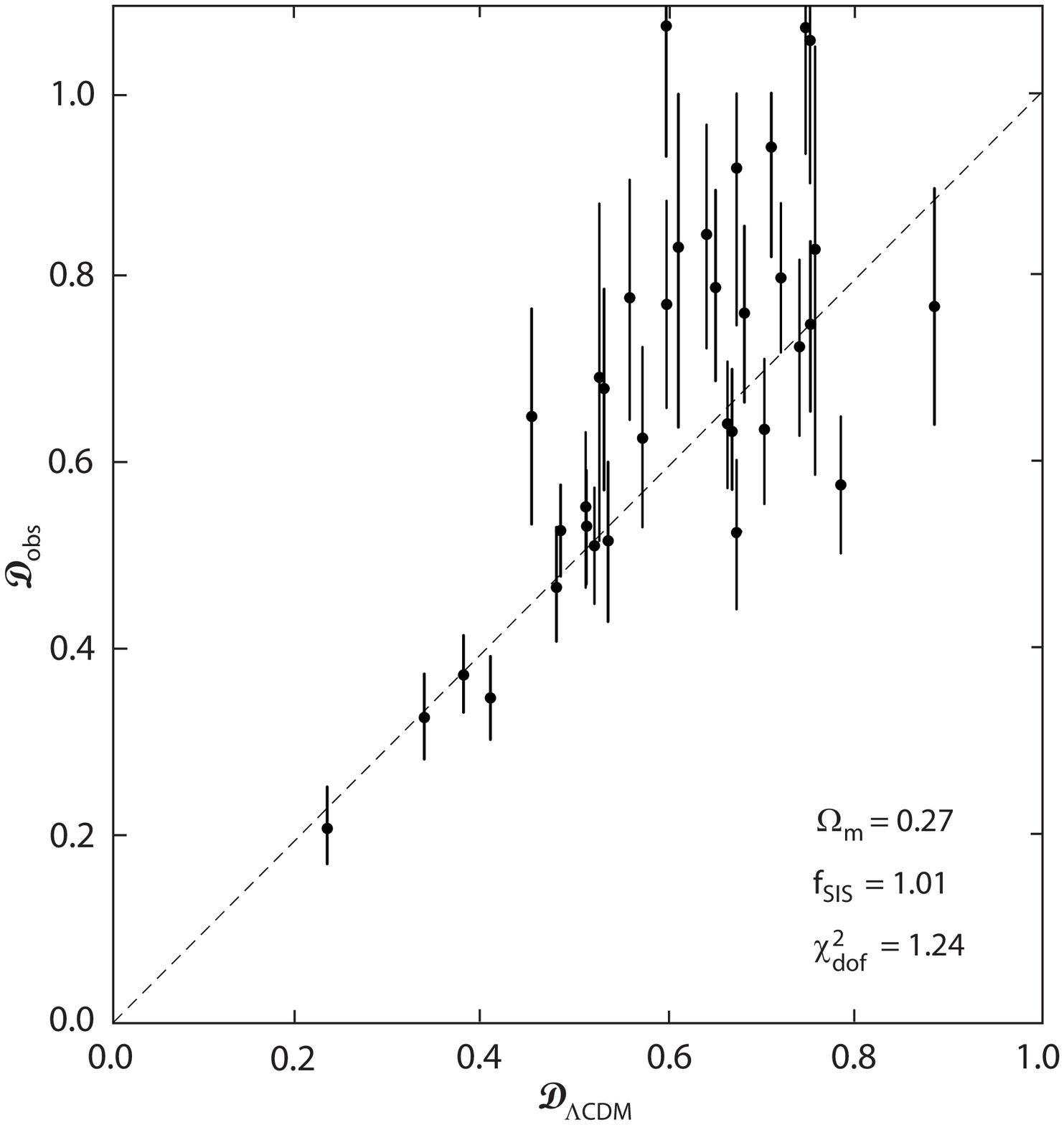}}
\vskip-0.2in
\caption{Same as figure~1, except now for the concordance $\Lambda$CDM
model, with $\Omega_{\rm m}=0.27$. The optimized value of $f_{\rm SIS}$
is $1.004$, and the reduced $\chi^2$ for this fit is $1.24$, with
$34-1=33$ degrees of freedom.}
\end{figure}

As we have already noted, the power to optimize model parameters,
such as $\Omega_{\rm m}$ in a multi-parameter context, is very
limited (Biesiada et al. 2010; Cao et al. 2012). This will become
quite apparent in our discussion below, where we compare the
quality of the fit for several variations of $\Lambda$CDM. For each
model we consider here, we will therefore optimize the fit using only
$f_{\rm SIS}$ as a free parameter, which we do by minimizing the
$\chi^2$ function
\begin{equation}
\chi^2(f_{\rm SIS})=\sum_i {\left(\mathcal{D}_{{\rm obs},i}[f_{\rm SIS}]-
\mathcal{D}_{{\rm th},i} \right)^2\over \sigma_{\mathcal{D},i}^2}\;,
\end{equation}
where the index $i$ runs over all lens systems in the sample, `th' stands
for either $\Lambda$CDM or $R_{\rm h}=ct$, and $\sigma_{\mathcal{D},i}^2$
is the variance of $\mathcal{D}_{{\rm obs},i}$ calculated from equation~(9).

\section{Discussion}
We have used the data shown in Table 1 to compare 3 variations of $\Lambda$CDM and
the $R_{\rm h}=ct$ Universe, though always for a flat universe ($k=0$) and
$w_{\rm de}=-1$. A summary of the results is provided in Tables 2 and 3 for the whole
sample (of 69 sources), and in Tables 4 and 5 for a reduced sample with
$\mathcal{D}_{\rm obs} <1$ only (63 systems). We note,
first of all, that the optimized value of $f_{\rm SIS}$ is very close to 1 in
every case, in complete agreement with earlier findings, e.g., by Treu et al. (2006),
and van de Ven et al. (2003). As such, we do not find any possible dependence of
the so-called bulge-halo `conspiracy' on the assumed cosmological model.

\begin{deluxetable}{lccc}
\tablewidth{277pt}
\tabletypesize{\footnotesize}
\tablecaption{Model Comparison for the Whole Sample}\tablenum{2}
\tablehead{Model&\colhead{$\qquad\Omega_{\rm m}\qquad$}&\colhead{$\qquad f_{\rm SIS}\qquad$}&
\colhead{$\qquad\chi^2_{\rm dof}\qquad$}
} \startdata

$R_{\rm h}=ct$ & .... &  1.023 &  1.22 \\
$\Lambda$CDM$^a$ &  0.20&  1.00 & 1.22 \\
{\rm Concordance}$^a$ &  0.27 &  1.01 &  1.24 \\
Einstein-de Sitter\qquad &  1.00 & 1.046 & 1.33 \\
\enddata
\tablenotetext{\null\hbox{$^a$}} {Assumes a cosmological constant with $w_\Lambda=-1$.
The concordance model is $\Lambda$CDM with $\Omega_{\rm m}=0.27$.
}
\end{deluxetable}

\begin{deluxetable}{lccc}
\tablewidth{277pt}
\tabletypesize{\footnotesize}
\tablecaption{Model Comparison for Two-image Sources}\tablenum{3}
\tablehead{Model&\colhead{$\qquad\Omega_{\rm m}\qquad$}&\colhead{$\qquad f_{\rm SIS}\qquad$}&
\colhead{$\qquad\chi^2_{\rm dof}\qquad$}
} \startdata

$R_{\rm h}=ct$ & .... &  1.033 &  1.27 \\
$\Lambda$CDM$^a$ &  0.20&  1.01& 1.28 \\
{\rm Concordance}$^a$ &  0.27 &  1.02 &  1.29 \\
Einstein-de Sitter\qquad &  1.00 & 1.059 & 1.33 \\
\enddata
\tablenotetext{\null\hbox{$^a$}} {Assumes a cosmological constant with $w_\Lambda=-1$.
The concordance model is $\Lambda$CDM with $\Omega_{\rm m}=0.27$.
}
\end{deluxetable}

\begin{deluxetable}{lccc}
\tablewidth{277pt}
\tabletypesize{\footnotesize}
\tablecaption{Model Comparison for all $\mathcal{D}_{\rm obs}<1$}\tablenum{4}
\tablehead{Model&\colhead{$\qquad\Omega_{\rm m}\qquad$}&\colhead{$\qquad f_{\rm SIS}\qquad$}&
\colhead{$\qquad\chi^2_{\rm dof}\qquad$}
} \startdata

$R_{\rm h}=ct$ & .... &  1.01 &  0.99 \\
$\Lambda$CDM$^a$ &  0.20&  0.99& 0.99 \\
{\rm Concordance}$^a$ &  0.27 &  1.00 &  1.00 \\
Einstein-de Sitter\qquad &  1.00 & 1.03 & 1.09 \\
\enddata
\tablenotetext{\null\hbox{$^a$}} {Assumes a cosmological constant with $w_\Lambda=-1$.
The concordance model is $\Lambda$CDM with $\Omega_{\rm m}=0.27$.
}
\end{deluxetable}

\begin{deluxetable}{lccc}
\tablewidth{277pt}
\tabletypesize{\footnotesize}
\tablecaption{Model Comparison for 2-images and $\mathcal{D}_{\rm obs}<1$}\tablenum{5}
\tablehead{Model&\colhead{$\qquad\Omega_{\rm m}\qquad$}&\colhead{$\qquad f_{\rm SIS}\qquad$}&
\colhead{$\qquad\chi^2_{\rm dof}\qquad$}
} \startdata

$R_{\rm h}=ct$ & .... &  1.02 &  0.92 \\
$\Lambda$CDM$^a$ &  0.20&  1.00& 0.92 \\
{\rm Concordance}$^a$ &  0.27 &  1.00 &  0.93 \\
Einstein-de Sitter\qquad &  1.00 & 1.05 & 0.99 \\
\enddata
\tablenotetext{\null\hbox{$^a$}} {Assumes a cosmological constant with $w_\Lambda=-1$.
The concordance model is $\Lambda$CDM with $\Omega_{\rm m}=0.27$.
}
\end{deluxetable}

We have compared the model fits using both the full sample of 69 entries in
Table 1 and, separately, using only the sub-sample of 34 2-image systems.
The quality of the fit, for every model we considered, is actually somewhat better
for the former. This may simply be a reflection of the fact that, though
technically an isothermal sphere should produce only 2 images, the other
possible effects described in \S~2 could be significant enough to result in
a considerable scatter about the average value $f_{\rm SIS}\sim 1$
for individual systems, that dwarfs all the other possible
sources of error in calculating $\mathcal{D}_{\rm obs}$. Indeed, Cao et
al. (2012) have argued that $\sigma_f$ could be as large as $\sim 20\%$,
and even though such a large scatter was not seen by Treu et al. (2006)
and van de Ven et al. (2003), they nonetheless did report an rms deviation
of at least $6-7\%$.

Given the universal result $f_{\rm SIS}\sim 1$, the entries
in column 6 of Table 1 are shown for only one value ($f_{\rm SIS}
=1.02$) of this fraction, corresponding to the optimized fit for the
$R_{\rm h}=ct$ model using only the 2-image lens systems (with
 $\mathcal{D}_{\rm obs} <1$). Also, column 9 shows the entries for 
$\mathcal{D}_{\Lambda{\rm CDM}}$
only for the concordance model, i.e., for $\Omega_{\rm m}=0.27$. These
values change somewhat for other choices of $\Omega_{\rm m}$, but not enough
to warrant showing all of them here.

Figures~1 and 2 demonstrate graphically how the observed values of $\mathcal{D}$
compare with those predicted by $R_{\rm h}=ct$ and the concordance model using
only the sub-sample of 2-image lens systems, but for all values of 
$\mathcal{D}_{\rm obs}$. The optimized value of
$f_{\rm SIS}$ is $1.02$ for the former, and $1.01$ for the latter,
and the reduced $\chi^2_{\rm dof}$ (with $34-1=33$ degrees of freedom) is quite
similar for these two cases, i.e., $1.22$ for the former versus $1.24$ for the
latter. It is quite evident from these figures that the scatter in
$\mathcal{D}_{\rm obs}$ about the theoretical curves (the straight
dashed lines in these plots) increases significantly as $D_A(z_l,z_s)
\rightarrow D_A(0,z_s)$. In other words, it appears that measuring
$\mathcal{D}$ becomes progressively less precise as the distance to the
gravitational lens becomes a smaller and smaller fraction of the distance
to the quasar source. This may simply have to do with the fact that $\theta_{\rm E}$
changes less and less for large values of $z_s/z_l$ so, for the same
error in the Einstein angle, one gets less precision in the
measurement of $\mathcal{D}_{\rm obs}/\mathcal{D}_{\rm th}$.

The principal results of this paper are summarized in Tables~2, 3, 4 and 5.
The first two show how well the 4 models considered here fit the complete
sample of 69 lens systems in Table 1 (for all values of  $\mathcal{D}_{\rm obs}$), 
whereas the latter two give the corresponding results for the  reduced
sample with  $\mathcal{D}_{\rm obs} <1$. Based
on the general trends emerging from these numbers, it is safe to draw
the following conclusions: (1) Even though the power of $\mathcal{D}$
to discriminate between different values of $\Omega_{\rm m}$ in $\Lambda$CDM
is quite limited, this analysis indicates that values larger than
$\Omega_{\rm m}=0.27$ probably don't work as well as those below it,
though the differences in $\chi^2_{\rm dof}$ are still too small
to draw any firm conclusions. And (2), the $R_{\rm h}=ct$ fits the
strong gravitational lens data at least as well as $\Lambda$CDM.
Still, the differences between $R_{\rm h}=ct$ and $\Lambda$CDM are
small enough that one cannot choose one model over the other based
solely on this analysis, using the current sample of strong
gravitational lens systems. The other tests we have completed thus far,
using, e.g., cosmic chronometers (Melia \& Maier 2013), high-$z$
quasars (Melia 2013), and gamma ray bursts (Wei et al. 2013), have
all resulted in a clear preference for $R_{\rm h}=ct$ over
$\Lambda$CDM using statistical model selection tools. The
analysis of the strong gravitational lensing data does not
result in a comparable outcome yet, though it too does not
provide any evidence that the standard model is a better
fit to these observations than $R_{\rm h}=ct$.

\section{Monte Carlo Simulations with a Mock Sample}
In order to provide a detailed quantitative assessment of what kind of strong 
lensing data are necessary to really distinguish the $R_{\rm h}=ct$ Universe 
from the standard $\Lambda$CDM model, we will here produce mock samples
of strong gravitational lenses based on the current measurement accuracy.
Several information criteria commonly used to differentiate between different 
cosmological models (see, e.g., Melia \& Maier 2013, and references cited 
therein) include the Akaike Information Criterion, ${\rm AIC}=\chi^{2}+2n$, 
where $n$ is the number of free parameters (Liddle 2007),
the Kullback Information Criterion, ${\rm KIC}=\chi^{2}+3n$ (Cavanaugh
2004), and the Bayes Information Criterion,
${\rm BIC}=\chi^{2}+(\ln N)n$, where $N$ is the number of data points
(Schwarz 1978). In the case of AIC, with ${\rm AIC}_\alpha$
characterizing model $\mathcal{M}_\alpha$,
the unnormalized confidence that this model is true is the Akaike
weight $\exp(-{\rm AIC}_\alpha/2)$. Model $\mathcal{M}_\alpha$ has likelihood
\begin{equation}
P(\mathcal{M}_\alpha)= \frac{\exp(-{\rm AIC}_\alpha/2)}
{\exp(-{\rm AIC}_1/2)+\exp(-{\rm AIC}_2/2)}
\end{equation}
of being the correct choice in this one-on-one comparison. Thus, the difference
$\Delta \rm AIC \equiv {\rm AIC}_2\nobreak-{\rm AIC}_1$ determines the extent
to which $\mathcal{M}_1$ is favoured over~$\mathcal{M}_2$. For Kullback
and Bayes, the likelihoods are defined analogously. In using the model selection tools,
the outcome $\Delta\equiv$ AIC$_1-$ AIC$_2$ (and analogously for KIC and BIC)
is judged `positive' in the range $\Delta=2-6$, `strong' for $\Delta=6-10$,
and `very strong' for $\Delta>10$.
In this section, we will estimate the sample size required to significantly strengthen the
evidence in favour of $R_{\rm h}=ct$ or $\Lambda$CDM, by conservatively
seeking an outcome even beyond $\Delta\simeq11.62$, i.e., we will see what is
required to produce a likelihood $\sim 99.7\%$ versus $\sim 0.3\%$,
corresponding to a $3\sigma$ confidence level.

We will consider two cases: one in which the background cosmology is
assumed to be $\Lambda$CDM, and a second in which it is $R_{\rm h}=ct$,
and we will attempt to estimate the number of strong gravitational lenses
required in each case in order to rule out the alternative (incorrect)
model at a $\sim 99.7\%$ confidence level. The synthetic strong gravitational lenses
are each characterized by a set of parameters denoted as ($z_{l}$, $z_{s}$,
$\sigma_{SIS}$, $\theta_{\rm E}$). We generate the synthetic sample using the
following procedure:

1. The simulations are carried out based on the current lens measurements.
We assign the lens redshift $z_{l}$ uniformly between $0.1$
and $1.1$, the source redshift $z_{s}$ uniformly between $1.5$ and $3.5$, and
the velocity dispersion $\sigma_{SIS}$ uniformly
between $100$ and $300$ km $\rm s^{-1}$,
as Paraficz \& Hjorth (2009) did in their simulations.

2. With the mock $z_{l}$, $z_{s}$ and $\sigma_{SIS}$, we first infer $\theta_{\rm E}$
from Equation~(7) corresponding either to the $R_{\rm h}=ct$ Universe (\S~5.1) or
a flat $\Lambda$CDM cosmology with $\Omega_{\rm m}=0.27$ (\S~5.2). We then assign a
deviation ($\Delta \theta_{\rm E}$) to the $\theta_{\rm E}$ value, i.e.,
we infer $\theta'_{\rm E}$ from a normal distribution whose center value is $\theta_{\rm E}$,
with a dispersion $\sigma=0.12\;\theta_{\rm E}$. The typical value of $\sigma$ is taken from
the current (observed) sample, which yields a mean and median deviation of $\sigma=0.15\;\theta_{\rm E}$
and $\sigma=0.11\;\theta_{\rm E}$, respectively. We constrain the mock sample to easily
detectable systems, so we include in the simulations only lenses with $\theta'_{\rm E}$
larger than $0.1$ arcsec.

\begin{figure}[h]
\centerline{\includegraphics[angle=0,scale=1.0]{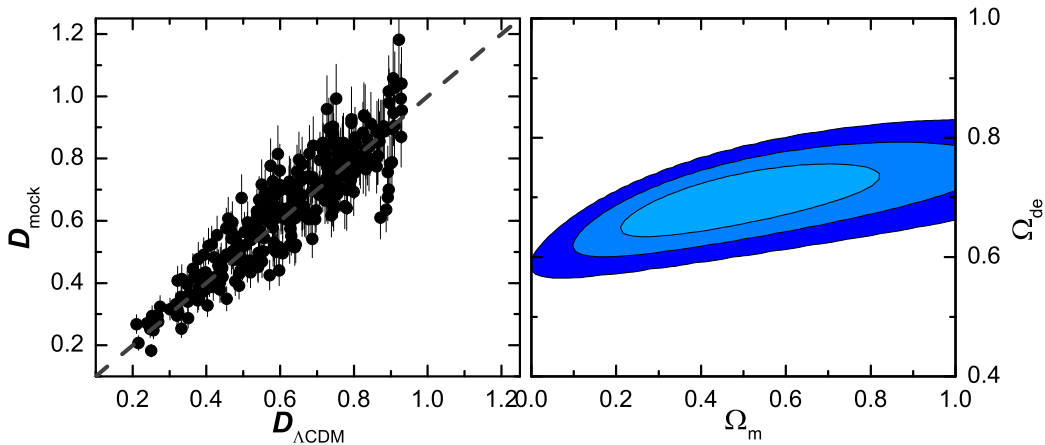}}
\vskip-0.2in
\caption{Left: ``Observed" values of $\mathcal{D}$ versus that predicted in the
best-fit $\Lambda$CDM model. A perfect fit would correspond to the dashed
diagonal line. The optimized values of $\Omega_{\rm m}$ and $\Omega_{\rm de}$
in this case are $0.45$ and $0.69$, respectively, and the reduced $\chi^2$ is $1.32$, with $300-2=298$ degrees
of freedom. Right: The $1\sigma-3\sigma$ contours corresponding to the parameters
$\Omega_{\rm m}$ and $\Omega_{\rm de}$ in the best-fit $\Lambda$CDM model, using the simulated sample with
300 lens systems, assuming $R_{\rm h}=ct$ as the background cosmology.}
\end{figure}

3. Assign observational errors to $\sigma_{SIS}$ and $\theta'_{\rm E}$.
Since both the observed errors $\sigma_{\sigma_{SIS}}$ and $\sigma_{\theta_{\rm E}}$
are about $5\%$ of $\sigma_{SIS}$ and $\theta'_{\rm E}$, we will assign the dispersions
$\sigma_{\sigma_{SIS}}=0.05\;\sigma_{SIS}$ and $\sigma_{\theta_{\rm E}}=0.05\;\theta'_{\rm E}$
to the synthetic sample.

This sequence of steps is repeated for each lens system in the sample, which
is enlarged until the likelihood criterion discussed above is reached. As with
the real 60-lens sample, we optimize the model fits by minimizing the $\chi^{2}$ function
in Equation~(12).

\subsection{Assuming $R_{\rm h}=ct$ as the Background Cosmology}
We have found that a sample of at least 300 strong gravitational lenses is required
in order to rule out $\Lambda$CDM at the $\sim 99.7 \%$ confidence level, if the 
background cosmology is in fact $R_{\rm h}=ct$.
The left-hand panel of Figure~3 show how the ``observed" values of $\mathcal{D}$ compare
with those predicted by the best-fit $\Lambda$CDM model using the simulated sample with
300 lens systems, assuming $R_{\rm h}=ct$ as the background cosmology. The
optimized parameters corresponding to the best-fit $\Lambda$CDM model for
these simulated data are displayed in the right-hand panel of Figure~3. To allow for the greatest
flexibility in this fit, we relax the assumption of flatness and allow
$\Omega_{\rm de}$ to be a free parameter along with $\Omega_{\rm m}$. The right-hand panel of Figure~3
shows the 2-D plots for the $1\sigma-3\sigma$ confidence regions for $\Omega_{\rm m}$
and $\Omega_{\rm de}$. The best-fit values for $\Lambda$CDM using the simulated sample with 300 lens systems
in the $R_{\rm h}=ct$ Universe are $\Omega_{\rm m}=0.45_{-0.24}^{+0.37}$ $(1\sigma)$ and
$\Omega_{\rm de}=0.69_{-0.06}^{+0.07}$ $(1\sigma)$, with a $\chi^2$ per degree of freedom of $\chi^2_{\rm dof}=394.04/298=1.32$.

\begin{figure}[hp]
\centerline{\hskip 0.5in\includegraphics[angle=0,scale=0.7]{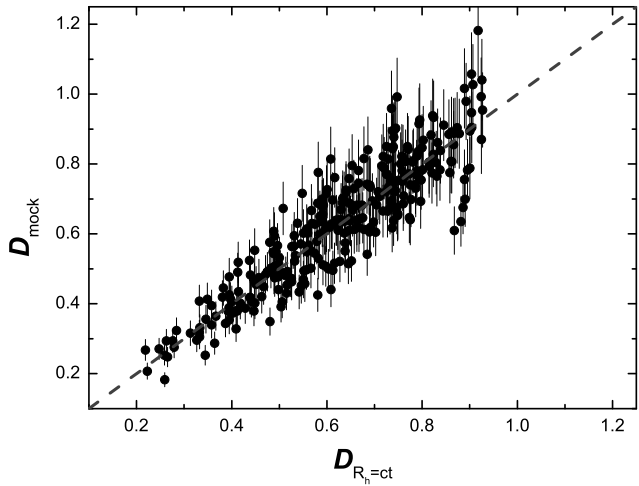}}
\vskip-0.2in
\caption{``Observed" values of $\mathcal{D}$ versus that predicted in the
$R_{\rm h}=ct$ universe, using a sample of 300 lens systems, simulated with $R_{\rm h}=ct$ as the
background cosmology. A perfect fit would correspond to the dashed diagonal line.}
\end{figure}

In Figure~4, we show how the ``observed" values of $\mathcal{D}$ compare
with those predicted in the $R_{\rm h}=ct$ universe. Note that there are no free parameters 
in $R_{\rm h}=ct$. With 300 degrees of freedom, the reduced $\chi^2$ is $\chi^2_{\rm dof}=
394.01/300=1.31$. 

Since the number $N$ of data points in the sample is now much greater than one, the
most appropriate information criterion to use is the BIC. The logarithmic penalty
in this model selection tool strongly suppresses overfitting if $N$ is large
(the situation we have here, which is deep in the asymptotic regime). With $N=300$,
our analysis of the simulated sample shows that the BIC would favour the $R_{\rm h}=ct$
Universe over $\Lambda$CDM by the aforementioned likelihood of $99.7\%$ versus only $0.3\%$
(i.e., the prescribed $3\sigma$ confidence limit).

\begin{figure}[hp]
\centerline{\includegraphics[angle=0,scale=1.0]{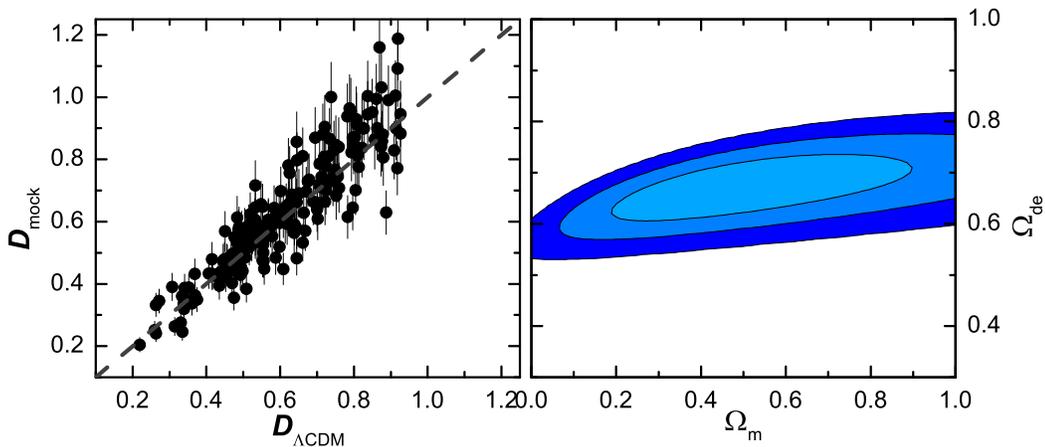}}
\vskip-0.2in
\caption{Same as Figure~3, except now with a flat $\Lambda$CDM as the (assumed)
background cosmology. The simulated model parameter was $\Omega_{\rm m}=0.27$.}
\end{figure}

\subsection{Assuming $\Lambda$CDM as the Background Cosmology}
In this case, we assume that the background cosmology is $\Lambda$CDM,
and seek the minimum sample size to rule out $R_{\rm h}=ct$ at the
$3\sigma$ confidence level. We have found that a minimum of 200 strong 
gravitational lenses are required to achieve this goal. To allow for the greatest flexibility
in the $\Lambda$CDM fit, here too we relax the assumption of flatness and
allow $\Omega_{\rm de}$ to be a free parameter along with $\Omega_{\rm m}$.
The left-hand panel of Figure~5 demonstrates how the ``observed" values of $\mathcal{D}$ compare
with those predicted by the best-fit $\Lambda$CDM model using the simulated sample with
200 lens systems, assuming $\Lambda$CDM as the background cosmology.
In the right-hand panel of Figure~5, we show 2-D plots of the $1\sigma-3\sigma$ confidence regions 
for $\Omega_{\rm m}$ and $\Omega_{\rm de}$. The best-fit values for $\Lambda$CDM using this simulated sample
with 200 lens systems are $\Omega_{\rm m}=0.47_{-0.28}^{+0.43}$ $(1\sigma)$ and 
$\Omega_{\rm de}=0.67_{-0.07}^{+0.07}$ $(1\sigma)$, with a $\chi^2$ per degree of freedom of $\chi^2_{\rm dof}=259.90/198=1.31$.

The ``observed" values of $\mathcal{D}$ compared with those predicted in the $R_{\rm h}=ct$ universe 
are shown in Figure~6. The dashed diagonal line denotes the perfect fit. 
With 200 degrees of freedom, the reduced $\chi^2$ is $\chi^2_{\rm dof}=281.86/200=1.41$. 
With $N=200$, our analysis of the simulated sample shows that in this case the BIC would favour $\Lambda$CDM
over $R_{\rm h}=ct$ by the aforementioned likelihood of $99.7\%$ versus only $0.3\%$
(i.e., the prescribed $3\sigma$ confidence limit).

\begin{figure}[hp]
\centerline{\hskip 0.5in\includegraphics[angle=0,scale=0.7]{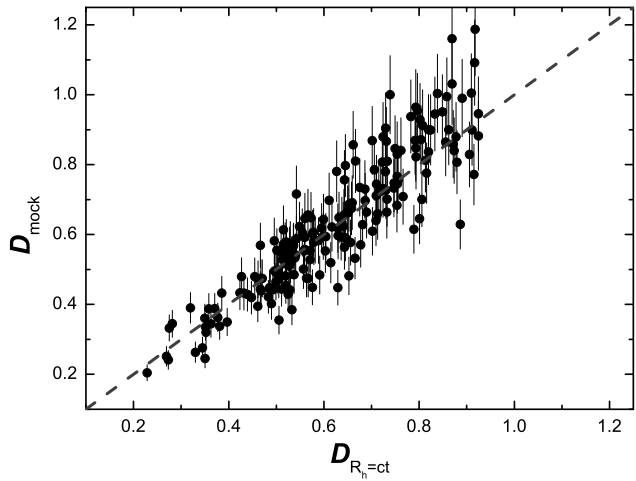}}
\vskip-0.2in
\caption{Same as Figure~4, except now with $\Lambda$CDM as the (assumed)
background cosmology.}
\end{figure}

\section{Conclusions}
The use of individual gravitational lenses to measure cosmological parameters
has been with us for over a decade now (see, e.g., Treu et al. 2006; Grillo et al. 2008;
Biesiada et al. 2010; and Biesiada et al. 2011) and the results, though less
precise than those from other kinds of data, have nonetheless been consistent with
the basic $\Lambda$CDM cosmology. Our principal goal in this paper has been to
carry out a comparative analysis of the available galaxy lens data using both $\Lambda$CDM
and the $R_{\rm h}=ct$ Universe, which may be thought of as $\Lambda$CDM with the
additional constraint $p=-\rho/3$ on its total equation of state. This analysis
has been motivated by other kinds of study showing that model selection tools
tend to favor $R_{\rm h}=ct$ over the standard model.

Insofar as the strong gravitational lenses are concerned, both $R_{\rm h}=ct$
and $\Lambda$CDM fit the data quite well. We have not found an inconsistency
between these results and those of previous studies using a variety of observations
at low and high redshifts. We may already be able to rule out values of $\Omega_{\rm m}$
much greater than the concordance value of $0.27$, but apparently not smaller than
this. Where things stand now is that gravitational lens data do not provide
conclusive evidence in favor of either model.

As much as we have learned about these lens systems, several sources of
uncertainty remain, including the need to properly model the mass distribution within
the lens and to better understand the source of the so-called bulge-halo conspiracy.
These errors appear to be more debilitating for lens systems with large values
of $z_s/z_l$, so a priority for future work ought to be the search for lens
systems with small distances between the lens and the source compared with
distances between the lens and observer. We have found that systems with
correspondingly small values of $\mathcal{D}_{\rm obs}$ provide significantly
greater precision in the measurement of cosmological parameters than those
with values approaching $1$ (see Figures~1 and 2).

Given the limitations of the current sample, we have also investigated how 
big the catalog of measured lensing galaxies has to be in order for us to rule 
out one (or more) of these models. We have considered
two synthetic samples with characteristics similar to those of the current observed
lens systems, one based on a $\Lambda$CDM background cosmology, the other on
$R_{\rm h}=ct$. From the analysis of these simulated lenses, we have
estimated that a sample of about 200 systems would be needed to rule out
$R_{\rm h}=ct$ at a $\sim 99.7\%$ confidence level if the real cosmology
were in fact $\Lambda$CDM, while a sample of at least 300 systems would
be needed to similarly rule out $\Lambda$CDM if the background cosmology
were instead $R_{\rm h}=ct$. The difference in required sample size
results from $\Lambda$CDM's greater flexibility in fitting the data, since
it has a larger number of free parameters.

Looking to the future, a convincing demonstration that $R_{\rm h}=ct$ is
the correct cosmology would provide sweeping new capabilities for
carrying out structural and evolutionary studies of lensing galaxies,
for the very simple reason that $\mathcal{D}$ in this cosmology is completely
independent of any model parameters, such as $H_0$ and $\Omega_{\rm m}$.
The quantity $\mathcal{D}_{\rm th}$ in this spacetime depends solely on
the observed values of $z_l$ and $z_s$ which, as we have noted in this paper,
are measured with much higher precision than any of the other lens-dependent
parameters. Imagine, therefore, the probative power of such measurements
on a determination of individual $f_{\rm SIS}$'s or, even better, on
providing the capability to probe the mass structure within these galaxies.

As of now, early-type galaxies appear to be well approximated by singular
isothermal ellipsoids. But this mass-density profile differs significantly
from cosmologically motivated ones (see, e.g., Navarro et al. 1997;
Moore et al. 1998), and also appears to require fine-tuning between the
distributions of baryonic and dark matter. This awkward situation begs
the question of how these structures formed in the first place. The
use of gravitational lensing within the $R_{\rm h}=ct$ framework may
finally break this deadlock and explain the origin of the bulge-halo
conspiracy.

\vskip-0.2in
\acknowledgments
We are grateful to the anonymous referee for providing a thoughtful review 
and for suggesting several improvements to the manuscript.
This work is partially supported by the National Basic Research Program (``973" Program) of China
(Grants 2014CB845800 and 2013CB834900), the National Natural Science Foundation of China
(grants Nos. 11322328,11373068, 11173064 and 11233008), the One-Hundred-Talents Program
and the Youth Innovation Promotion Association, and the Strategic Priority Research Program
``The Emergence of Cosmological Structures'' (Grant No. XDB09000000) of the Chinese Academy 
of Sciences, and the Natural Science Foundation of Jiangsu Province. F.M. is also grateful to 
Amherst College for its support through a John Woodruff Simpson Lectureship, and to Purple 
Mountain Observatory in Nanjing, China, for its hospitality while this work was being carried out.
This work was partially supported by grant 2012T1J0011 from The Chinese Academy of
Sciences Visiting Professorships for Senior International Scientists, and grant
GDJ20120491013 from the Chinese State Administration of Foreign Experts Affairs.


\begin{thebibliography}

\bibitem[Bartelmann \& Schneider (1999)]{1999A&A...345...17B} Bartelmann, M. \& Schneider, P.\ 1999,
A\&A, 345, 17

\bibitem[Biesiada et al. (2010)]{2010MNRAS.406.1055B} Biesiada, M., Pi\'orkowska, A. \& Malec, B.\
2010, \mnras, 406, 1055

\bibitem[Biesiada et al. (2011)]{2011RAA....11..641B} Biesiada, M., Malec, B. \& Pi\'orkowska, A.\ 2011,
RAA, 11, 641

\bibitem[Bolton et al. (2005)]{2005ApJ...624L..21B} Bolton, A., Burles, S. M., Koopmans, L.V.E.,
Treu, T. \& Moustakas, L. M.\ 2005, \apjl, 624, L21

\bibitem[Bolton et al. (2006)]{2006ApJ...638..703B} Bolton, A., Burles, S. M., Koopmans, L.V.E.,
Treu, T. \& Moustakas, L. M.\ 2006, \apj, 638, 703

\bibitem[Bolton et al. (2008)]{2008ApJ...682..964B} Bolton, A. S., Burles, S., Koopmans, L.V.E., Treu, T.,
Gavazzi, R., Moustakas, L. A., Wayth, R. \& Schlegel, D. J.\ 2008, \apj, 682, 964

\bibitem[Cao et al. (2012)]{2012JCAP...03..016C} Cao, S., Pan, Y., Biesiada, M.,
Godlowski, W. \& Zhu, Z.-H.\ 2012, JCAP, issue 3, id. 16

\bibitem[Cavanaugh (2004)]{Cav04} Cavanaugh, J. E. 2004, Aust. N.~Z. J. Stat., 46, 257

\bibitem[Fassnacht et al. (1996)]{1996ApJ...460L.103F} Fassnacht, C. D., Womble, D. S.,
Neugebauer, G., Browne, I.W.A., Readhead, A.C.S., Matthews, K. \& Pearson, T. J.\ 1996,
\apjl, 460, L103

\bibitem[Futamase \& Yoshida (2001)]{2001PThPh.105..887F} Futamase, T. \& Yoshida, S.\ 2001,
Progress Theor. Phys., 105, 887

\bibitem[Grillo et al. (2008)]{2008A&A...477..397G} Grillo, C., Lombardi, M. \& Bertin, G. \
2008, A\&A, 477, 397

\bibitem[Hewett \& Wild (2010)]{2010MNRAS.405.2302H} Hewett, P. C. \& Wild, V.\ 2010,
MNRAS, 405, 2302

\bibitem[Huchra et al. (1985)]{1985AJ.....90..691H} Huchra, J., Gorenstein, M., Kent, S.,
Shapiro, I., Smith, G., Horine, E. \& Perley, R.\ 1985, \aj, 90,691

\bibitem[Koopmans \& Treu (2002)]{2002ApJ...568L...5K} Koopmans, L.V.E. \& Treu, T.\
2002, \apj, 568, 5

\bibitem[Koopmans \& Treu (2003)]{2003ApJ...583..606K} Koopmans, L.V.E. \& Treu, T.\
2003, \apj, 583, 606

\bibitem[Koopmans et al. (2006)]{2006ApJ...649..599K} Koopmans, L.V.E., Treu, T., Bolton, A. S.,
Burles, S. \& Moustakas, L.A.\ 2006, \apj, 649, 599

\bibitem[Koopmans et al. (2009)]{2009ApJ...703L..51K} Koopmans, L.V.E. et al.\ 2009, \apjl,
703, L51

\bibitem[Kormann et al. (1994)]{1994A&A...284..285K} Kormann, R., Schneider, P. \& Bartelmann, M.\ 1994,
A\&A, 284, 285

\bibitem[Leh\'ar et al. (1993)]{1993AJ....105..847L} Leh\'ar, J., Langston, G. I.,
Silber, A., Lawrence, C. R. \& Burke, B. F.\ 1993, \aj, 105, 847

\bibitem[Liddle (2007)]{2007MNRAS.377L..74L} Liddle, A. R. 2007, MNRAS, 377, L74

\bibitem[Melia(2007)]{2007MNRAS.382.1917M} Melia, F.\ 2007, \mnras, 382, 1917

\bibitem[Melia(2013)]{2013ApJ...764...72M} Melia, F.\ 2013, \apj, 764, 72

\bibitem[Melia \& Maier(2013)]{2013arXiv1304.1802M} Melia, F., \& Maier, R.~S.\ 2013, MNRAS, 432, 2669

\bibitem[Melia \& Shevchuk(2012)]{2012MNRAS.419.2579M} Melia, F., \& Shevchuk, A.~S.~H.\ 2012, \mnras, 419, 2579

\bibitem[Moore et al. (1998)]{1998ApJ...499L...5M} Moore, S. M., Governato, F., Quinn, T., Stadel, J.
\& Lake, G.\ 1998, \apjl, 499, L5

\bibitem[Narayan \& Bartelmann (1996)]{1996astro.ph..6001N} Narayan, R. \& Bartelmann, M.\ 1996,
arXiv:astro-ph/9606001

\bibitem[Navarro, Frenk \& White (1997)]{1997ApJ...490..493N} Navarro, J., Frenk, C. S.
\& White, S.D.M.\ 1997, \apj, 490, 493

\bibitem[Newton et al. (2011)]{2011AAS...21734705N} Newton, E. R., Marshall, P. J. \& Treu, T.,
SLACS Collaboration\ 2011, \apj, 734, 104

\bibitem[Paraficz \& Hjorth 2009]{2009A&A...507L..49P} Paraficz, D. \& Hjorth, J.\ 2009,
A\&A Letters, 507, L49

\bibitem[Perlmutter et al. (1999)]{1999ApJ...517..565P} Perlmutter, S. et al. 1999, ApJ, 517, 565

\bibitem[Ratnatunga et al. (1999)]{1999AJ....117.2010R} Ratnatunga, K. U., Griffiths, R. E.
\& Ostrander, E. J.\ 1999, AJ, 117, 2010

\bibitem[Refregier (2003)]{2003ARA&A..41..645R} Refregier, A.\ 2003, ARAA, 41, 645

\bibitem[Riess et al. (1998)]{1998AJ....116.1009R} Riess, A. G. et al. 1998, AJ, 116, 1009

\bibitem[Schneider et al. (1992)]{1992grle.book.....S} Schneider, P., Ehlers, J. \& Falco, E. E.\ 1992,
Gravitational Lenses (Springer Verlag, Berlin)

\bibitem[Schwarz (1978)]{Sch78} Schwarz, G. 1978, Ann. Statist., 6, 461

\bibitem[Suyu et al. (2013)]{2013ApJ...766...70S} Suyu, S. H. et al.\ 2013, \apj, 766, 70

\bibitem[Tonry (1998)]{1998AJ....115....1T} Tonry, J. L.\ 1998, \aj, 115, 1

\bibitem[Treu \& Koopmans (2002)]{2002ApJ...575...87T} Treu, T. \& Koopmans, L.V.E.\ 2002,
\apj, 575, 87

\bibitem[Treu \& Koopmans (2003)]{2003MNRAS.343L..29T} Treu, T. \& Koopmans, L.V.E.\ 2003,
\mnras, 343, 29

\bibitem[Treu \& Koopmans (2004)]{2004ApJ...611..739T} Treu, T. \& Koopmans, L.V.E.\ 2004,
\apj, 611, 739

\bibitem[Treu et al. (2006)]{2006ApJ...640..662T} Treu, T., Koopmans, L.V.E., Bolton, A. S.,
Burles, S. \& Moustakas, L. A.\ 2006, \apj, 640, 662

\bibitem[van de Ven et al. (2003)]{2003MNRAS.344..924V} van de Ven, G., van Dokkum, P. G.
\& Franx, M.\ 2003, MNRAS, 344, 924

\bibitem[Wei et al. (2013)]{2013ApJ...772...43W} Wei, J.-J., Wu, X.-F. \& Melia, F.\ 2013,
\apj, 772, id.43

\bibitem[Wei et al. (2014a)]{2014ApJ...788..190W} Wei, J.-J., Wu, X.-F. \& Melia F.\ 2014a,
\apj, 788, id.190

\bibitem[Wei et al. (2014b)]{AJTypeIa} Wei, J.-J., Wu, X.-F. \& Melia, F. 2014b, AJ, submitted

\bibitem[White \& Davis (1996)]{1996AAS...189.4104W} White, R. E. \& Davis, D. S.\ 1996, BAAS,
28, 1323

\bibitem[Young et al. (1980)]{1980ApJ...241..507Y} Young, P., Gunn, J. E., Kristian, J.,
Oke, J. B. \& Westphal, J. A.\ 1980, \apj, 241, 507

\end{thebibliography}
\end{document}